\documentstyle[aps]{revtex}

\begin{document}
\draft

\title{Phase-field model with reduced interface diffuseness}

\author{Seong Gyoon Kim}
\address{Department of Materials Science and Engineering\\Kunsan National University\\
Kunsan 573-701, Korea}

\author{Won Tae Kim}
\address{Department of Physics\\Chongju University\\Chongju 360-764, Korea}

\author{Toshio Suzuki}
\address{Department of Materials Engineering\\The University of Tokyo\\
Tokyo 113, Japan}

\date{\today}
\maketitle
\begin{abstract}
  We minimized the interface diffuseness  in the phase-field models  by introducing the 
parabolic double-well potential  and localizing  the solute  redistribution (or  latent heat 
release) into a   narrow region  within  the  phase-field interface.   In spite  of  the 
parabolic potential with cusps, highly  localized  solute redistribution  and discontinuous 
diffusivity function  adopted  in this   model, it  works remarkably   well in  numerical 
computations. The  computations on  dendritic  solidification  of  an one-sided  system 
yield quantitatively the same results  with the  anti-trapping model  [A. Karma,   Phys. 
Rev. Lett.   87, 115701  (2001)], indicating  the anomalous   interfacial effects   can be  
effectively suppressed. This approach can be easily extended to the multi-components or 
multi-phases system.

\end{abstract}

\section{INTRODUCTION}  
  Solidification processes in  materials produce complex  microstructural patterns in  the 
interface morphology and composition distribution  profile. The prediction and  control of  
solidification microstructures are important  because the properties  of the materials  are 
usually dictated  by the  microstructure. However  the computational  prediction  of the 
microstructure has been one of the formidable free boundary  problems. Traditional front 
tracking methods solving  the equations of  motion of  the sharp interface  and keeping 
tracking the interface position in every  time steps have been used,  but appeared to be 
not suitable especially when the geometry  is complex or in 3D.  The phase-field model 
(PFM) has emerged as  an effective tool  in describing the  complex pattern  evolution 
\cite{1}. In this approach, the phase field $\phi$ is introduced, varying from one value 
in solid to another value in  the liquid across a spacially diffuse  interface with a width 
$2 \xi$. The equations governing  the phase  field and  diffusion  field in PFM  follow  
naturally from the definition of the free  energy functional of the  system and requiring 
it  to decrease  monotonically in time.  In particular,  Karma and  Rappels'  asymptotic 
analysis \cite{2} at the thin interface limit ($2 \xi \ll D / V$, $D$: thermal diffusivity,  
$V$: interface  velocity) improved  greatly the  computational efficiency  and made  the  
PFM to be  a quantitative computational  tool under  the real  experimental  conditions, 
even though their approach was restricted  to  the case  of pure materials' solidification  
with equal thermal diffusivity  and heat capacity in solid and liquid phases. 

  Several PFMs for  alloys \cite{3,4,5}  also have  been  proposed. To  use the   alloy 
PFMs as  the computational  tools  that can  yield quantitative   results under realistic 
conditions, it is highly desirable  to determine the parameters  in the PFMs at  the thin 
interface limit,  not  at the  sharp  interface limit.    As shown by   Almgren \cite{6}, 
however, Karma  and   Rappel's asymptotic  analysis  can not   be  straightforwardly 
extended  to  the alloy  PFMs   because the  unequal solute   diffusivities and solute 
partitioning between solid and liquid  give rise to anomalous interface effects;  chemical 
potential jump,  surface  stretching and   surface diffusion. If  the heat  capacities and 
thermal diffusivities are different in solid  and liquid, the PFMs for pure materials suffer 
from the same  anomalous interface  effects. Recently, Karma  \cite{7} proposed  a new 
PFM  for  alloy  solidification, which  is based on i) introduction  of an  anti-trapping 
flux  in the  diffusion equation  and ii)  assumption  of negligible solute  diffusivity in  
solid ($D_S$), compared with that in liquid ($D_L$). The anti-trapping  flux could   be  
manipulated to  not only  cancel out the solute trapping  current across the interface at 
low  interface   velocity,  but  also   eliminate  all  the   anomalous  interface  effects 
simultaneously, together with proper choices  for some auxiliarly functions.  Furthermore 
the assumption $D_S \ll D_L$, as  in the our previous approach  of 1-D thin interface 
analysis \cite{5}, make the  steady diffusion equation to be integrated  without unknown 
constant,  which enable us to find the  relationship between the phase-field mobility and 
the kinetic coefficient at  the thin interface limit. 

   In this study we propose a simple method for getting rid  of the anomalous interface 
effects in the PFMs,  which can be  straightforwardly extended to  the multi-phases or 
multi-components system. The anomalous interface  effects in PFMs originate  from the 
diffuseness of the interface  in the diffusion equation,  irrespective of the diffuseness  in 
the phase-field equation; they all vanish when the sharp interface  limit is taken for the 
diffusion equation. Furthermore, the interface  width in PFMs can be  defined differently 
each other  for  diffusion  field   and for  phase  field;   $2  \xi_d$ and $2  \xi_p$, 
respectively. Our approach thus  is based on the  decoupling the  interface widths  for 
both fields within the thin interface limit  and pushing  the interface width for diffusion 
field to  be minimized. Mathematically this corresponds to the condition $D_L  /V \gg 
2 \xi_p \gg 2  \xi_d  \rightarrow 0  $, in contrast  with  the previous thin  interface 
limit $D_L  /V \gg2  \xi_d \sim  2  \xi_p$ \cite{2}.  In  spirit,   present approach  is 
somewhat  similar to  the  pseudo-front   tracking  method \cite{8} proposed  recently, 
where the diffusion equation  is  solved on the  sharp interface and  the  curvature for  
the  Gibbs-Thomson  equation  is   obtained    from the   diffuse    interface. In    
followings, we describe   our approach in detail and then present   computation  results 
on  the  dendritic growth. 

\section{MODEL}
  We start from  finding the  governing equations unified   for both the  pure  thermal 
case  and the  pure  solutal case. With   definitions  $\phi=-1$ at liquid,  $\phi=1$ at 
solid and $-1 <  \phi <1$ at  the interface region,  the phase-field equation  \cite{2} is 
given by 
\begin{equation}
\tau \partial_t \phi = W^2 \nabla^2 \phi -\partial_\phi [ f(\phi) + \lambda u g(\phi) ],  
\label{1}
\end{equation}
where $\lambda = a_1 W / d_0 $, $d_0$ is the capillary length, $f(\phi)$  is the double 
well potential with  the  minima at  $\phi=\pm 1$,  $g(\phi)$  is an odd   function of 
$\phi$ with  $g'( \pm  1) =0$,  $u$  is  the dimensionless   field defined   as $u  = 
(T-T_m )/   ( \Delta H_m /C_p^L )$ for  pure  materials and  $u = (c_L  -c_L^e ) /  
( c_L^e -c_S^e  )$ for  alloys, where $T_m$ is  the  melting point,  $\Delta  H_m$ is  
the latent heat,  $C_p^S$ and  $C_p^L$ are  the specific   heats of  solid and  liquid, 
respectively and  $c_L^e$  and $c_S^e$ are the  equilibrium  composition of  solid and 
liquid  at a   given temperature, respectively. When the constant $a_1$ is taken as 
\begin{equation}
a_1 = {1 \over {g(1)-g(-1) }} \int_{-1}^1 \sqrt {2  f (\phi_0 )} d \phi_0. 
\label{2}
\end{equation}
Eq. (\ref{1}) recover  the Gibbs-Thomson  equation $u=-  d_0 k$ at  equilibrium state, 
where $k$ is the  curvature of the  interface. For solidification  of pure materials,  the 
diffusion equation can be written as 
\begin{equation}
\partial_t H  = D_L  \nabla \cdot  q(\phi )  [C_p^S {{1+h(\phi)}  \over  2}  + C_p^L
{{1-h(\phi)} \over 2} ] \nabla T,
\label{3}
\end{equation}
where the local enthalpy   density $H$ is  given by $H=  H_S [1+h(\phi)] /  2 + H_L 
[1-h(\phi)] / 2$, $H_S$ and  $H_L$ are the enthalpy  densities of solid and  liquid  as 
functions  of temperature,  respectively, $h(\phi)$  is an  odd function of  
$h( \pm 1)= \pm 1$, $C_p$ is the specific heat given by $C_p =C_p^S [1+h(\phi)] / 2 + 
C_p^L [1-h(\phi)]  /  2$, the    diffusivity function  $q(\phi)$ is    defined  to satisfy 
$q(-1)=1$ and   $q(1)=D_S /D_L$,   where $D_S$  and  $D_L$   are   the  thermal 
diffusivities   in   solid and   liquid, respectively.   For   isothermal   solidification of 
binary alloys,  the solute diffusion equation \cite{4,5} can be written as 
\begin{equation}
\partial_t c  = D_L   \nabla \cdot  q(\phi) [   ( {{1+h(\phi)}  \over 2}   \nabla  c_S  + 
{{1-h(\phi) } \over 2}  \nabla c_L ],
\label{4}
\end{equation}
where the local composition  (mole fraction) $c$ is given by $c=c_S [1+h(\phi)]/2 + c_L 
[1-h(\phi)] /  2$, $D_L$   is the solute   diffusivity  in liquid  and the   corresponding 
$q(\phi)$ of  solute  diffusivity to   that of   the thermal   diffusivity.  Although  the 
chemical potential that can  be defined  everywhere may  be  used  instead of   $c_S$ 
and  $c_L$ as  in \cite{5,7}, it is  more convenient   to use  Eq. (\ref{4}) directly,  as 
long  as $c_L$  and $c_S$  are defined  at the  whole  space of  the system.   With 
$H_S =  H_S^m +C_p^S  (T-T_m )$, $H_L =   H_L^m +C_p^L (T-T_m  )$  and the 
usual  condition  $c_S  /  c_L   =c_S^e /  c_L^e$ \cite{4,5},   the  diffusion equations 
(\ref{3}) and (\ref{4}) can be written  as an unified form; 
\begin{equation}
{\partial \over {\partial t}} [  u A(\phi)  - {1 \over  2} h(\phi)  ]  = D_L  \nabla \cdot 
q(\phi)A(\phi) \nabla u, 
\label{5}
\end{equation}
where $A(\phi)=[1 + k -( 1- k  )h(\phi)]/2$, and $k= C_p^S / C_p^L$  for solidification 
of pure materials and $k=c_S^e /  c_L^e$ for isothermal solidification of  binary  alloys. 
Note that in  fact  the final forms  of the governing equations (\ref{1}) and (\ref{5}) in 
this study are fundamentally  identical with those   in the previous  studies \cite{4,5,7}, 
except  the anti-trapping term in the diffusion equation of Ref. \cite{7}.    

  The interface width $2 \xi_p$ of phase field is determined by the form of the  double 
well potential $f(\phi)$ in  Eq. (\ref{1}). In  this study a  parabolic potential $f(\phi)= | 
1-\phi^2 |/2$ is adopted, instead  of the traditional form  $(1-\phi^2 )^2$. The parabolic 
potential is not new  one, but has been  used in PFMs \cite{9,10}.  Comparing with the 
fourth order potential,  the parabolic potential gives several benefits as  follows; i)  The 
equilibrium phase-field  profile is given  by  $\phi_0  = -\sin  (x/  W )$.  Outside  of 
the interface width $2 \xi_p = \pi W$, the  phase state becomes completely  either solid 
or liquid,  which  is  contrast  to  long  smearing  of  phase  field  into  solid   and  
liquid phases in  the PFMs  with  fourth  order  potential. This clear  cut  of interface 
region suppresses the   spurious  interaction  between  interfaces   during computation 
on  the  system with    complex   interface  geometry;   For  example,  when two   
solid/liquid interfaces  are   approaching  each  other,  the  two   interfaces remains   
intact until the  distance  between  two  interface  becomes less   than  $\pi W$.   
However, in case  of the fourth  order potential,  two approaching   interfaces  interact  
each other   with spurious  long-range attraction  and then   merges easily   into an 
interface, due to  the  overlapping  of  smearing phase  fields at  liquid  between two 
solids.  ii)  Below we will  introduce  a highly localized $h(\phi)$  form  to  minimize 
the anomalous  interface effects.  However coupling the $h(\phi)$ with the fourth order 
potential makes the phase-field profile  to be less stable  in computation with increased 
grid size,  compared with  the coupling with the   parabolic potential.   iii) Importantly,  
the parabolic   potential is   more convenient  in extending  the phase-field   model to 
multiphase-fields case   than  the  fourth order    potential, because  combination  of   
the parabolic potentials  gives  local maxima   at the triple   junction \cite{10}.  Indeed 
the multi-phases-field model with the  parabolic potentials is  extensively used recently  
for quantitative  computation of    the microstructure  under   the  real experimental    
conditions \cite{11}.
              
  The interface  width  $2 \xi_d$  for  diffusion field  is  determined by  the  function 
$h(\phi)$ in Eq. (\ref{5}). Here we adopt a simple form;
\begin{equation}
h(\phi) = 
\cases{
-1 ,  & \text{for $\phi < -\phi^* $;} \cr
\phi / \phi^* , &\text{for $-\phi^* < \phi< +\phi^* $;} \cr  
+1 , &\text{for $\phi>+\phi^* $.} \cr }
\label{6}
\end{equation}
This form of $h(\phi)$ represents that  solute redistribution or latent heat release  takes 
place in a  localized region  $- \phi^*  <\phi <  +\phi^*$ within  the interfacial  region 
($-1<\phi<1$) of the phase field. The $\phi^*$  represents the  degree   of localization,  
 the smaller  the  value  the   more localization of  solute redistribution. The $\phi^*$  
value defines a new interface width $2  \xi_d$ with  the  relationship $ 2\xi_d =   2W 
\sin^{-1} \phi^*$, and the  width is independent  of  the interface width  $2  \xi_p$ of 
phase  field.  If we take $\phi^* =1$, then   it follows    $h(\phi) =\phi$,   which has 
been widely    used in  previous  approaches  \cite{2,4}.    All   the   anomalous   
interface   effects   become smaller    with decreasing $2\xi_d$    and  completely 
disappear  at  the limit  $2  \xi_d \rightarrow 0$.  In reality of numerical  computation, 
however, vanishing $2 \xi_d$ can  not be taken as  long   as a finite  mesh  size  is  
used  in computations: Consider   1D interface moving steadily    in computation with  
mesh size $\Delta   x$.  If  we  set $2  \xi_d$  (over where  $\phi$  changes from   
$-\phi^*$ to $+\phi^*$) to be  smaller   than $\Delta x$, there   can be   the moment 
when the phase-field  values on   all  the grid  points are outside  the range  $-\phi^* 
< \phi  < +\phi^*$.  During  this  moment,  the system  loses the   existence of  the 
interface region  for diffusion field. Even though  this phenomena does  not lead to  the 
violation  of the  mass or   energy  conservation, it plays  the  role of a very   large 
noise source in  computations. Thus  the best choice of the  width $2 \xi_d$ appears to 
be $2 \xi_d = \Delta x$.

   As long as we can not take the  limit $2 \xi_d \rightarrow 0$, we should  take care 
of the  anomalous interface  effects. The  conditions for  vanishing anomalous  interface 
effects \cite{6,7} are given by 
\begin{equation}
\int_{-\xi_d}^0 [1-h(\phi_0 )] dx  = \int_0^{+{\xi}_d}  [1+h(\phi_0 )] dx, 
\label{7}
\end{equation}
for the interface stretching, 
\begin{equation}
\int_{-{\xi}_d}^0 {{ 1-h(\phi_0 )}  \over { q(\phi_0  ) A(\phi_0 )}} dx   = 
\int_0^{+{\xi}_d}  [2- {{ 1-h(\phi_0 )} \over {q(\phi_0 ) A(\phi_0 )}} ] dx, 
\label{8}
\end{equation}
for the diffusion potential (chemical potential, concentration or temperature) jump and
\begin{equation}
\int_{-{\xi}_d}^0 q(\phi_0 )A(\phi_0  ) dx = \int_0^{+{\xi}_d} 
[1- q(\phi_0 ) A(\phi_0 )] dx, 
\label{9}
\end{equation}
for  the  interface  diffusion. $\phi_0$ is the equilibrium phase-field  profile given by $ 
\phi_0 = -\sin  (x/ W )$ for  the parabolic potential in this study.   It should be noted 
that for  the last condition (\ref{9})  we  assumed $q(\phi_0 ) =D_S  /D_L \rightarrow 
0$ at $x<  -\xi_d$.  The  first  condition   (\ref{7})  is satisfied   with  the  form 
(\ref{6})  for  $h(\phi_0 )$.  The second  condition also   can be  satisfied by  simply  
putting $q(\phi_0 )  A(\phi_0 )=1$  at $x > -\xi_d$.   The last condition  however  is  
violated, as  long as the  conditions (\ref{7}) and (\ref{8}) are satisfied. How large  the 
anomalous   interface diffusion  effect is   in this  case? With $q(\phi_0  ) A(\phi_0 )  
=0$ at $x< -\xi_d$ and  $q(\phi_0 ) A(\phi_0 )  =1$  at   $x >  -\xi_d$, the   mass  
balance  condition \cite{7}  at the thin  interface limit  becomes 
\begin{equation}
V= -D_L [ \partial_n u^+ - \xi_d  \partial_s^2 u ],
\label{10}
\end{equation}
where $V$ is the interface  velocity, $u^+$ is the diffusion  field at the liquid  side  of 
the  interface. The  second  term in  the right hand  side of   Eq.  (\ref{10}) is  the 
anomalous diffusion  flux along the   arc length  of  the interface. This flux originates 
from putting  $q(\phi_0 ) A(\phi_0 )  =1$ at $-\xi_d < x <  0$ within the solid  side 
of the  interfacial region,  and the  anomalous  interface diffusion occurs within the half 
interface width $\xi_d$. With the choice $2 \xi_d  = \Delta x$ as explained before, this 
width corresponds to $\Delta x/2$. Whether  or not  the anomalous  interface diffusion 
within  the half   mesh size affects much the dynamics  of the interface  is  dependent 
on  the computational systems  and conditions.  The effect   can  be significant  when 
the curvature  gradient inducing the interface  diffusion is   very large. Such  situation 
may be  met at the  tip region during computations of dendritic growth  with   a 
large mesh size,  which will  be  tested in the   later part of  this study. At  the thin 
interface  limit and under the  conditions $D_S  \ll D_L$  and $q(\phi_0 ) A(\phi_0  ) 
=1$ at $x >  -\xi_d$,  following Ref.  \cite{2,5},  we  find   the   velocity-dependent 
Gibbs-Thomson   equation $u=-d_0 k -\beta V$ with the kinetic coefficient 
\begin{equation}
\beta = {{ a_1 \tau } \over {\lambda  W}}[ 1 - a_2 \lambda  {W^2 \over {\tau 
D_L}} ],
\label{11}
\end{equation}
where the constant $a_2$ is given by 
\begin{equation}
a_2 = {1 \over a_1} \int_{-1}^{+1}  {{g(\phi_0 )-g(-1)} \over {g(1)-g(-1) 
}} \cdot {{ 1-h(\phi_0 )} \over {\sqrt{2f(\phi_0 )} }} d \phi_0 , 
\label{12}
\end{equation}
which appears  to  be  identical with   that in  Ref.  \cite{2,7} if  the  constant $K$ 
therein is integrated  partially, and also  can be  found in  our  previous study  \cite{5} 
by putting $q(\phi_0 ) A(\phi_0 )   =1$ at the  interfacial   region.  As    pointed out  
in Ref. \cite{7},  Eq.  (\ref{11}) and (\ref{12})  remain  unchanged  in the   symmetric 
case  also  where $q(\phi)=1$ and $A(\phi)=1$ in Eq. (\ref{5}). 

\section{COMPUTATION}  
   In present model, we  adopted the parabolic  potential $f(\phi)= |  1-\phi^2 |/2$ with 
cusps at $\phi=\pm 1$, the   step-like function $h(\phi)$ defined  by  Eq. (\ref{6}) and  
the discontinuous  function  $q(\phi)$   satisfying  $q(\phi   )  =0$   at  $\phi   >  
\phi^*$ (equivalently, at  $x<  -\xi_d$) and   $q(\phi )  A(\phi  ) =1$  at  $\phi  < 
\phi^*$.  The combination  of  such   functions in   the  model   looks  apparently   
unsuitable  in  view of the continuum spirit of PFMs.  However the model works 
remarkably well in numerical computations  to be shown.  

  We simulated two cases. The first simulation was done on the dendritic growth in the 
symmetric case with $q(\phi)=1$ ($D_S  = D_L$) and $A(\phi)=1$ ($k=1$).  We tested 
the effect of the interface  width $2 \xi_d$ on the  accuracy in tip velocity, as  well as 
the performance of the parabolic potential.  The  second simulation   was done on   the 
one-sided  case with $D_S =0$  and $k <1$. We  tested the convergence with $d_0  / 
W$ and the self-consistency in  tip composition, and compared   the tip  velocity with 
that  from Karma's  model \cite{7}. We  solved the diffusion equation (\ref{5}) and the 
well-known anisotropic form \cite{2} of  Eq. (\ref{1}) by  the typical explicit  finite 
difference scheme.  The four fold   anisotropy in the interface   energy was introduced 
by putting $W(\theta) = W [1+ \epsilon_4 \cos 4  \theta ]$, where $\theta$ is the angle 
between the direction  normal to the  interface and $x$-axis.  The phase-field mobility 
was determined at  the vanishing kinetic coefficient  condition in Eq. (\ref{11});
\begin{equation}
\tau (\theta)  = a_2 \lambda  W(\theta)^2 /D_L . 
\label{13}
\end{equation}
The functions and  parameters commonly  used in both  simulations are  as followings; 
$f(\phi)= | 1-\phi^2 |/2$, $g(\phi)   = (3/2) (\phi -   \phi^3 /3 )$, $h(\phi)$  given  by 
Eq. (\ref{6}), $a_1  =\pi/4$, $a_2   =0.2637$ and  $0.4150$  for  $\phi^*  = 0.2$  and 
$0.95$, respectively, $W=1$,  $D_L =1$  and the  mesh size   $\Delta x   =0.4$. Note 
that  the interface width for the phase field   then  is  given by $2 \xi_p  =  \pi W = 
7.85 \Delta x$,  whereas the  interface  width  $ 2\xi_d  =  2W \sin^{-1} \phi^*$ for 
diffusion  field is given by   $1.00 \Delta  x$  and $6.27   \Delta x$  for  $\phi^*   = 
0.2$ and $0.95$, respectively. The initial undercooling (supersaturation) of the  melt was 
$u_0 = -  0.55$. The solidification  started from  the quarter circle  solid seed  of  the 
radius $r_0 =  40 d_0  $ at  one  corner  of the   system. The   initial  fields  before 
solidification were $u  =0$, $\phi=1$ at $r<r_0$ and $u =u_0 -u_0 \exp [ -d_0 (r-r_0 ) 
/(4 \Delta x) ]$, $\phi=-1$ at $r>r_0$.

   Two points  must be  explained for  the numerical  computations.  The first  is the 
stability problem related with the parabolic  potential. At the edges  $x=\pm \xi_p =\pm  
\pi W/2$ of the interface during  computation, the phase-field values can  be $\phi> 1$ 
or $\phi< -1$ and then oscillate around  $\phi=\pm 1$  as computation proceeds, which 
is due to $\partial_{\phi} f(\pm 1) \neq 0$  and  the cusps  at $\phi  =\pm 1$ in   the 
parabolic  potential. Those  oscillations propagate  into the   both the  bulk solid   and 
liquid region  to make  the interface unstable.  This instability   could  be avoided  by  
simply putting  $\phi=1$  if $\phi>1$ and  $\phi=-1$  if $\phi < - 1$ on  all grids in  
every time steps.  Even with such   control, however,  the interface became   unstable 
when  we took  $\phi^*  =1$   in Eq. (\ref{6}),  that  is  $h(\phi) =\phi$, which   is  
related with   the anomalous  release  of   latent  heat   (or solute  redistribution) at  
the  edges  of the  interface, following the artificial  cutting of   the phase-field values.  
For the  computational stability,  thus the adoption  of the parabolic potential  need  to  
be combined with  the localization  of $h(\phi)$, as $\phi^* <1$   in Eq. (\ref{6}). The 
second point  is  the   concentration   oscillation in  the  solid  of     the one-sided  
system,  which  originates  from   the combination of   the highly  localized $h(\phi)$  
and the discontinuous $q(\phi)$. After jumping from the  interfacial state with $-\phi^* 
<\phi <\phi^*$ to the solid  state  with  $\phi> \phi^*$  at  a given  time step,   the 
concentration of  the grid    is frozen. Thus  the  frozen concentration of the  solid is 
dependent  on the  composition of  the   grid  at  the last  time  step  remaining  as  
the interfacial state.   As the   result,  the frozen  composition of the    solid oscillates  
with solidification  length   and  the   oscillation   amplitude becomes   larger  with 
decreasing number of  grids  in $-\xi_d  <x< \xi_d$.  At the  dendrite   tip of    the  
one-sided   system, the   oscillation  amplitude   in  most computations with $2 \xi_d  
= \Delta x$ was about $10  \%$ of  the composition  shift  due  to   Gibbs-Thomson 
effect. This composition oscillation could be removed by putting  a small margin $\delta 
\phi$ in the definition  of $q(\phi)$;  $q(\phi) =1$  at $\phi <   \phi^*  +\delta  \phi$  
and  $q(\phi)=0$  otherwise.  The   minimum margin $\delta \phi_{\rm{min}}$ required  
to  remove   the  oscillation corresponds    to just   the phase-field   change  during 
one  time  step  in  a grid   with $\phi=\phi^*$.  In most computations of    present  
study,   $\delta \phi_{\rm{min}}$   was  less   than $0.01$.   Therefore we adopted 
$\delta  \phi  =0.01$  in  all   computations on    the one-sided system,  with which  
the  oscillation amplitude  in computations  with  $2  \xi_d  = \Delta  x$ could   be 
suppressed  within $0.1   \%$ of   the  composition    shift  due  to  Gibbs-Thomson 
effect. 
 
   In the dendrite growth  in undercooled pure  melt (symmetric system), the  effect of 
the localization  of  latent  heat   release was   tested. We  took  $k=1$, $\epsilon_4 
=0.05$ and $d_0 /W$ =0.277 on  300 x 1200 grid system.  In Fig.\ \ref{fig1}, the scaled 
tip velocity $Vd_0  /D_L$ was plotted  with the  scaled  time $tD_L /d_0^2$  for two 
different $\phi^*$ values  in Eq. (\ref{6}).  The thin line  and dotted line are the results 
with $\phi^* = 0.2$  and $\phi^*  =0.95$ from  this  study, respectively, the  thick line  
from the standard PFM,  and the horizontal line  is the  steady tip  velocity from   the 
Green function  method. For  the  standard  PFM, we used the usual  forms  $f(\phi)= 
(1-\phi^2 )^2 /4$, $ g(\phi) =  \phi -2 \phi^3 /3 + \phi^5 /5$,  $h(\phi)= \phi$, and then 
$a_1 =0.8839$   from Eq.  (\ref{2})   and $a_2  =0.6267$  from  Eq.  (\ref{12}).  As 
mentioned before,  the case  with $\phi^*   =1$ could not   be  tested  because   the 
combination of  the   parabolic  potential  with   the  corresponding  $h(\phi) = \phi$ 
makes the phase-field profile unstable. As can be seen in Fig.\ \ref{fig1}, two   extreme 
interface widths of $2 \xi_d = \Delta  x$ ($\phi^* =0.2$) and $2 \xi_d  \simeq 2 \xi_p$ 
($\phi^* =0.95$) for the diffusion field yield  the similar tip velocity  change  with time. 
In  particular, the tip velocity vs time for $\phi^* =0.95$ appeared to be nearly identical 
with that from the  standard  PFM $h(\phi)=\phi$.  The final scaled  tip  velocities  at 
steady state were  $0.0172$,  $0.0169$, $0.168$ for  $\phi^* =0.2$, $\phi^* =  0.95$ and 
the standard PFM, respectively, which  are very close to    the velocity  $0.0170$ from 
the Green function  method. Thus it   turns out that  the combination of  the parabolic 
potential  and the highly localized $h(\phi)$ works as well as the standard PFM.

   Next we present the computation results  for one-sided  system. Fig.\ \ref{fig2} 
shows the variation of  scaled tip  velocity  $Vd_0 /D_L$ with the  scaled time $tD_L 
/d_0^2$, where we took $k=0.15$ and $\epsilon_4  = 0.02$. The thin line and the dotted 
line are the  results  for $d_0  /W =0.277$ on  300 x 1200  grid system and  $d_0 /W 
=0.554$ on 600 x 2000 grid  system, with the same $\phi^*  =0.2$. The  dashed line is 
the result  for $d_0 /W  =0.277$ and  $\phi^* =0.95$  on 300   x 1200  grid   system. 
For comparison, the computational   results obtained  from  the  anti-trapping  current 
model \cite{7}  was  included as the   thick line   in Fig.\ \ref{fig2}.  Following   the 
notations  in this  study, his model   can  be    written as  the  phase-field  equation  
(\ref{1})  with $f(\phi)=   (1-\phi^2 )^2    /4$, $g(\phi)    =  \phi  -2   \phi^3  /3  
+\phi^5   /5$, $h(\phi)=\phi$  and the diffusion equation  
\begin{equation}
{\partial \over {\partial t}} [ u A(\phi) - {1 \over 2} h(\phi)  ]  
= {D_L \over 2} \nabla  \cdot (1-\phi) \nabla u
+ {W  \over {2 \sqrt{2}}}   \nabla \cdot   [1+(1-k)u] {{\nabla \phi} 
\over {| \nabla \phi |}} \phi_t .
\label{14}
\end{equation}
The phase-field mobility at  the vanishing kinetic coefficient  condition is given  by Eq. 
(\ref{13}), where  $a_1 =0.8839$ and $a_2 =0.6267$. As shown in Fig.\ \ref{fig2}, the tip 
velocity  variations with  time remain unchanged  for twice change in $d_0 /W$  when 
$2 \xi_d = \Delta x$ (or $\phi^* = 0.2$) was adopted. If the surface diffusion could not 
be negligible,  then its effect and  so the tip velocity would be   dependent on the grid 
size or   $d_0 /W$.   This convergence   thus indicates  that the   anomalous surface 
diffusion was  negligible even at  $d_0 /W = 0.277$.   The tip velocities for $\phi^*  = 
0.2$ are   in close  agreement with  that  (thick  line  in  Fig.\  \ref{fig2})  from the  
anti-trapping model \cite{7}. We also measured the  tip radius variation with the scaled 
time, following  the  method suggested in Ref. \cite{2}. The  tip radii for $\phi^* =0.2$ 
were within $5 \%$  error, compared  with  that  computed from   the   anti-trapping  
model.  These good agreement in  tip velocity  and tip   radius may  be seen  as  the  
evidence that  the  anomalous interface  phenomena  were  effectively suppressed   in 
both approaches.  We increased  the interface width   for the diffusion  field  over six 
times from  $2 \xi_d   = \Delta  x$ ($  \phi^* =0.2$)  to  $2  \xi_d =6.27   \Delta x$ 
($\phi^* =0.95$), while keeping $d_0 /W = 0.277$. The tip velocity then was  decreased 
significantly as seen from  the dashed line in Fig.\ \ref{fig2}, whereas  the tip   radius 
was increased  by  $40  \%$. Considering that   the surface diffusion current decreases 
the  curvature gradient of the dendrite tip,   these must  be the  manifestation of   the 
significant  anomalous surface diffusion effect.  Fig.\ \ref{fig3}   shows a  steady-state 
composition   profile across the interface along the growth axis  of  the   dendrite   in  
the   one-sided   alloy system  with $\epsilon_4  = 0.02$,  $k=0.15$, $d_0 /W  =0.277$ 
and $\phi^*    =0.2$. The  scaled solid   composition for   plane-front  interface  at    
equilibrium  state  is   $c_S^e /  c_L^e  =0.15$.  Note that   in Fig.\  \ref{fig3}  the  
solute redistribution  around the   interface occurs within  one grid  width only  due to  
the  localization  of $h(\phi)$. The  inserted magnified figure shows  the comparison of 
the measured  solid composition (filled  circles) with  prediction from   Gibbs-Thomson 
equation  (thick  horizontal  line),  $u= - d_0 / \rho$ or  equivalently  $c_S /c_L^e =k  
[1-(1-k )  d_0 / \rho   ]$, where  $\rho$ is  the computed  tip radius.      Excellent 
agreement   within  3  $\%$   error   can   be   noticed between the  computation 
results and the prediction.    

\section{SUMMARY}
   In summary we  minimized the  interface diffuseness  in the phase-field  models by 
introducing the parabolic double-well potential and localizing the solute redistribution (or 
latent heat release) into a narrow  region within a phase-field  interface. In spite of the 
parabolic potential with  cusps, highly  localized solute  redistribution and  discontinuous 
diffusivity in  this model,  it works  remarkably  well in  numerical computations.  The 
computations on dendritic solidification  of an one-sided system  yield quantitatively the 
same results  with Karma's  anti-trapping model,  indicating the   anomalous interfacial 
effects can  be effectively  minimized.  This approach  can be   easily extended to  the 
multi-components or multi-phases system. This  approach also is useful  in suppressing 
the anomalous   interaction between   interfaces in   the closely-spaced  multi-particles 
systems, e.g. liquid-phase sintering.

\begin{figure}
\caption{ Effect of $h(\phi)$ width on the scaled  tip velocity variations with the scaled  
time. The computational and material parameters were $k=1$, $u_0 =-0.55$,  $\epsilon_4  
= 0.05$, $W =1$, $D_L =D_S  =1$, $\Delta x / W =0.4  $ and the phase-field mobility 
$\tau$ was determined at the vanishing kinetics condition. }
\label{fig1}
\end{figure} 

\begin{figure}
\caption{Scaled tip velocity variations   with the scaled time  for one-sided alloy model 
with $D_S   =0$. The  computational and   material parameters  were  $k=0.15$, $u_0 
=-0.55$, $\epsilon_4   = 0.02$,  $W =1$,  $D_L =1$,  $\Delta  x /  W =0.4$  and the 
phase-field mobility $\tau$ was determined at the vanishing kinetics condition.}
\label{fig2}
\end{figure} 

\begin{figure}
\caption{The composition  profile across   the interface along  the  growth axis  of the 
dendrite and  comparison of   the measured   tip composition    with Gibbs-Thomson  
equation along the  growth axis  for  one-sided  alloy model   with $D_S =0$.   The 
computational and  material parameters  were $k=0.15$,  $u_0  =-0.55$, $\epsilon_4  = 
0.02$,  $d_0  /W =0.277$,   $W =1$, $D_L  =1$,  $\Delta  x  / W   =0.4$ and  the 
phase-field mobility $\tau$ was determined at the vanishing kinetics condition.}
\label{fig3}
\end{figure} 

\end{document}